# Creating A Coefficient of Change in the Built Environment After a Natural Disaster

**Karla Saldana Ochoa***
School of Architecture, SHARE Lab
University of Florida
1480 Inner Rd, Gainesville, FL 32611
`ksaldanaochoa@ufl.edu`

## Abstract

This study proposes a novel method to assess damages in the built environment using a deep learning workflow to quantify it. Thanks to an automated crawler, aerial images from before and after a natural disaster of 50 epicenters worldwide were obtained from Google Earth, generating a 10,000 aerial image database with a spatial resolution of 2 m per pixel. The study utilizes the algorithm Seg-Net to perform semantic segmentation of the built environment from the satellite images in both instances (prior and post-natural disasters). For image segmentation, Seg-Net is one of the most popular and general CNN architectures. The Seg-Net algorithm used reached an accuracy of 92% in the segmentation. After the segmentation, we compared the disparity between both cases represented as a percentage of change. Such coefficient of change represents the damage numerically an urban environment had to quantify the overall damage in the built environment. Such an index can give the government an estimate of the number of affected households and perhaps the extent of housing damage.

## 1 Introduction

In the housing sector, an operation that guarantees situational awareness is damage assessments in the built environment. One data modality that enables efficient and rapid damage assessment of the urban environment, whether in buildings, infrastructure, or green areas, is satellite imagery. In the literature, several researchers have proposed methodologies to analyze satellite images to create damage assessment ([1]; [2]; [3]; [4]; [5]; [6]; and [7]). In damage assessments, the most explored branch of research is Change Detection after a natural disaster. Change Detection is the process of identifying the difference in a building's state before and after a disaster using pre and post-disaster images [8]. Appendix 1 shows different methods, databases, and algorithms used to perform this task.

## 2 Application Context

The Change Detection result informs and guides several decisions in disaster response because it provides situational awareness after a natural disaster. In particular, such results influence decisions that have to do with the urban and housing sector. According to Duyne Barenstein et al. (2010), Change Detection in the built environment provides critical evidence for establishing a reconstruction policy. Rapid Change Detection on the built environment can give the government an estimate of the number of affected households and perhaps the extent of housing damage. Therefore, this study proposes a method that allows capturing a coefficient of change that categorize changes after a natural disaster from 50 natural disasters from the last ten years; such prototype aims to demonstrate that such methodology is helpful to be employed after a natural disaster, as a tool to quantify the damage and assign early financial help.

---

*www.ai-share-lab.com)

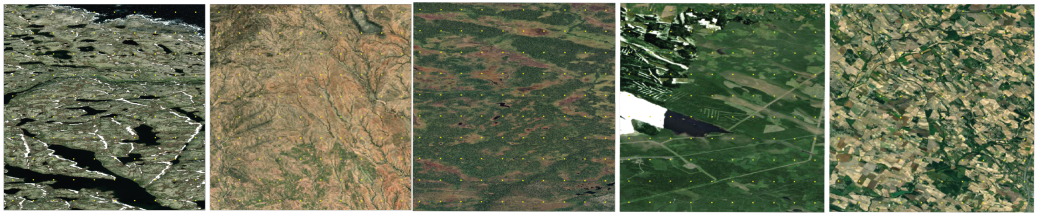

Figure 1: Sample of grids defined for crawling satellite image data.

## 3 Methodology

The methodology is divided into two main tasks: a)collection of satellite imagery data and b) creating a coefficient of change. The last task is further subdivided into image transformation, building Detection, and quantifying the coefficient of change.

### 3.1 Collection of Satellite Image

The present research focuses on the change after a natural disaster; the satellite images needed for this analysis must correspond to two specific dates before and after the natural disaster for each of the 50 natural disasters selected (Appendix 2 describes the characteristics of each disaster). It is always challenging to find 'good quality' satellite images when searching for a specific date. Especially when focusing on getting open access data. After carefully checking various free databases, we found that few data were available for many required dates. However, one database analyzed has satellite imagery available for all case studies. This database is Google Earth. This platform allows one to browse a timeline for specific geo-coordinates. That is to say, that one can find satellite images of the same place for specific dates.

To collect satellite images pre and post disaster from Google Earth, we first defined the dates selected for crawling, which had a range of 2 years before and after the disaster (from the 50 natural disasters). Then we set up a crawler to extract the satellite images from a list of geographic coordinates. These coordinates were defined by a geo-grid covering 10 km2 (Figure 01) with a separation of 100 meters, resulting in 100 geo-coordinates per epicenter. To start the crawling process, we set up an elevation level to 1500m in the Google Earth configuration to have the exact spatial resolution of 2m per pixel. After crawling, we collected a total of 10000 satellite images pre and post disaster covering 150,000 km.

When crawling the satellite images, we considered an overlap between them to help in the registration process. For the registration process, we used a Geometric Transformation method that only focused on translation. For such a transformation, we used the "Find Geometric Transform" algorithm (a Wolfram Mathematica function) that takes a corresponding pair of images and finds a geometric transformation that aligns positions specified by the key points in the first image and those that correspond with the second image. This returned a matrix that described the alignment error together with the transformation function. Such a matrix was then used on a "Linear Fractional Transform" algorithm (a Wolfram Mathematica function), which eventually registered these two images. Such a process was done for the 100 patches per epicenter. Figure 02 shows a subsample of the satellite images for each epicenter.

### 3.2 Creating a coefficient of change using Seg-net

The creation of a coefficient of change follows three steps: 1. Image transformation to make the before and after images comparable. 2. Building Detection and 3. The quantification of the change.

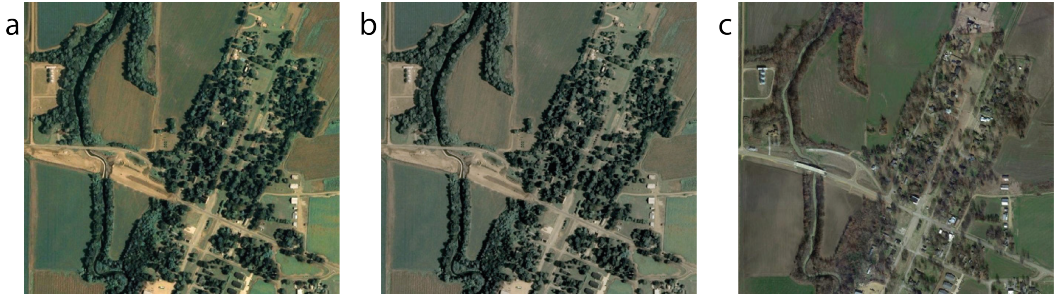

Figure 2: a) image before original, b) image before with the histogram transform, c) image after.

### 3.2.1 Image Transformation

Collecting images of the same place from two different dates may result in different resolutions since they may not come from the same satellite or may have been captured at different times or in different weather conditions. In order to identify the damage after a natural disaster, it is necessary to normalize the two images so that they can be compared. There are several methods for performing this normalization: shape analysis, histogram transformation, brightness value comparison, or image differencing by thresholding the color difference. For this experiment, we will use the histogram transformation (HT) method [9]. This method takes one of the two images as a reference, in our case, the image after the natural disaster, and modifies the pixel values of the image to be transformed, in our case the pre-disaster image, so that its histogram has almost the same distribution as the reference image.

After transforming all the pre-disaster images, we must ensure that the two images cover the same area. To achieve this, we applied an algorithm called Feature Tracking [10]. This algorithm compared the two images (before and after) to find corresponding key points. By applying this algorithm to the images, we obtained a list of common points between each pair of images. Then for each pair of images, we identified the key point closest to the center of the after-disaster image and assigned a radius of 8km around it. This radius ensures that the two images that previously covered an area of 100km2 now covered 64km2 of the same area. Although the images were collected with the exact geographic coordinates, there were many times when the same area was not obtained due to issues concerning how the satellite image was registered. After performing this normalization process, we obtained two images of the same place (the epicenter of the natural disaster) that captured the same amount of space and had similar RGB representations. The two methods used were coded in Wolfram Mathematica 12.2 using built-in functions within the software. Figure 03 shows an example of this process.

### 3.2.2 Building Detection

For the task of building detection, we used a pixel-based semantic segmentation method. Semantic segmentation is a process that takes an RGB image as input and produces an image of equal size colored by pixels based on semantic labels as output. All pixels with the same label were colored identically [11]. For this task, we trained a CNN model called Seg-Net. In Seg-Net, each module consists of three to five convolutional layers, with each one followed by one batch normalization layer. At the end of each module, there was one pooling layer and one activation layer. Then, the compressed images were fed into a set of hierarchical up-sampling modules. Each module started with an up-sampling layer and was followed by several convolutional layers, with the last one being followed by an activation layer. The pooling layer and the up-sampling layer in the modules of the same hierarchy share their pooling indices. The general architecture of the model is illustrated in Figure 04.

This model was trained with the ISPRS commission II/4 dataset[12]. These data were manually classified into six land cover classes: impervious surfaces, buildings, low vegetation, trees, cars, and background. These training images were partitioned in a 256x256x3 format and fed into a pre-trained Seg-Net, via a process known as transfer learning. After being trained on the ISPRS commission II/4 dataset, the newly trained model, which we call Seg-Net-Sat, learned the characteristics of the new dataset, which are different from the original training data. After training, we validated the Seg-Net-Sat model, arriving at a prediction accuracy of 92%. Afterward, we proceeded to predict the segmentation to the whole database of 100 satellite images collected before and after the disaster. As in the ISPRS dataset, we also partitioned the satellite images in 256x256x3 format to input them on the Seg-Net-Sat algorithm. After the segmentation of all the identified elements, only the building layer was extracted, which can be seen in Figure 05 as a black and white image. With this method, we obtained two black and white images per disaster describing the built environment, which permitted us to calculate the difference between them (Figure 05).

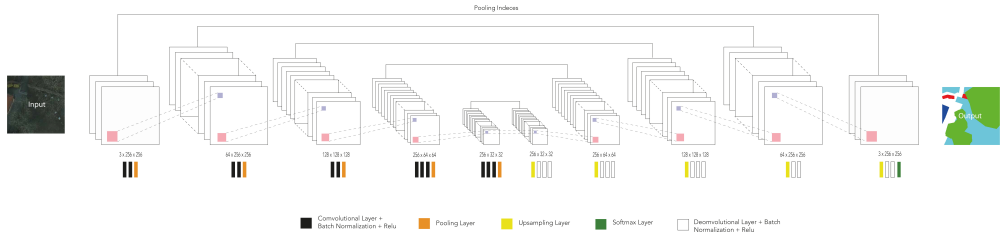

Figure 3: The architecture of the Seg-Net model where the disposition of the convolutional, pooling, up-sampling, softmax and normalization layers are explained.

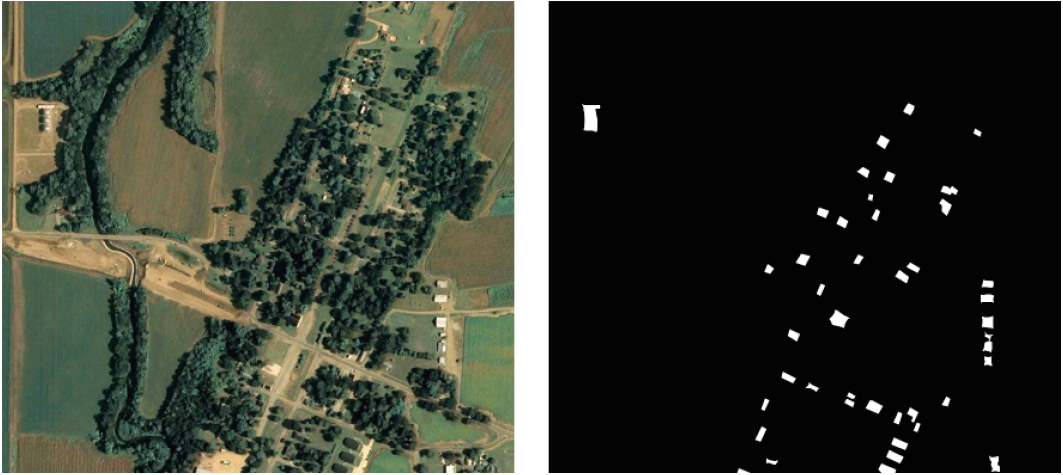

Figure 4: Satellite image and the corresponding building detection.

### 3.2.3 Quantification of change

To quantify the change in the built environment, we use a method called Object-based image analysis[13] (OBIA). This method focused on objects and not on a pixel-to-pixel comparison. By only identifying the built environment, we could make comparisons between objects. The segmented images presented the buildings that were found as white objects against a black background. These segmented buildings can be compared by counting the number of buildings per image (before and after). After obtaining the two results, we can calculate the difference between them, e.g., Disaster X, before-image had 1000 buildings, and Disaster X, after-image had 955 buildings the difference between these two images resulted in 45 missing buildings. After comparing the 50 pairs of images, it was possible to identify the most extreme case in terms of difference (missing buildings). This was set as the highest coefficient of change (1). The former defines a range of 1 as maximum change and 0 as minimum change. This range of change was assigned to each epicenter depending on the difference found between the two states. With this last step, we obtained a coefficient of change of the built environment after a natural catastrophe. Figure 06 shows the process for one disaster.

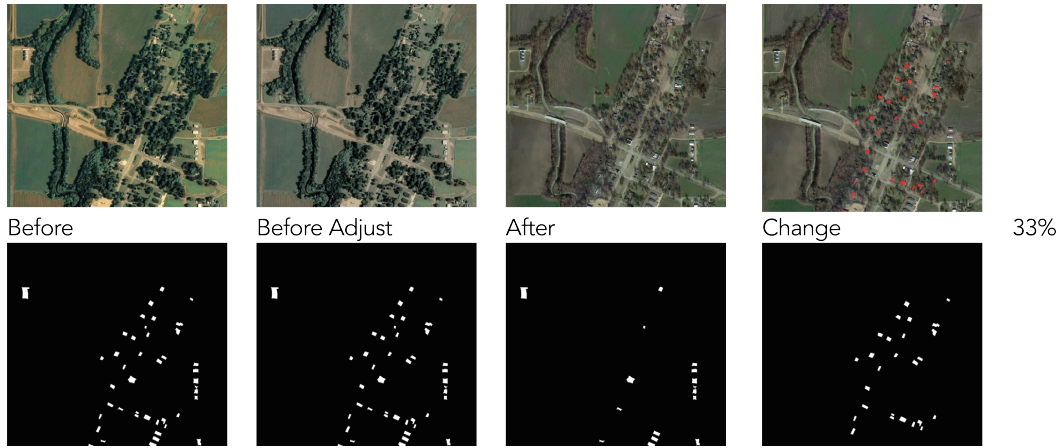

Figure 5: Change Detection Process.

## 4 Conclusion and Discussion

A limitation encountered in the experiment has to do with the satellite images necessary to create a coefficient of change. Some of the images collected did not have good resolution, meaning that the segmentation was sometimes inaccurate. However, the segmentation was consistent, so this inconsistency appeared throughout the database and became a constant. However, if the resolution of the satellite images improves, we will have better results. In other words, the calculated coefficient of change will be more accurate and reflect the current situation of the built environment in more detail.

A novel aspect of this experiment was its work with a collection of natural disasters. The experiment does not concentrate on isolated events but tries to create a space to compare among the various disasters considered. Another novelty was working with freely available data. The emphasis on using accessible databases is sincere because we would like to replicate this research with a larger number of case studies. Therefore, if the research's impact is to be scaled up, a large number of resources would have to be used, which in many cases would not be available, especially in developing countries. In this experiment, 50 natural disasters were considered, but future research will seek to expand this collection to have more answers that will serve as suggestions for responding to an ongoing natural disaster.

In future research, we will use such a coefficient of change in the built environment to cross-connected with financial data to respond to such disasters to find correlations that can inform the future response to natural disasters because we acknowledge that having information about damage to buildings is not sufficient for disaster response. Hence relating such findings (coefficient of change) with additional data, such as investment needed for recovery, will provide valuable information for various operations in disaster response. However, this relationship is rarely considered when allocating resources for disaster response because decisions are made in isolation without a clear framework for analyzing the capacity (from previous experiences) and determining the resources allocated for disaster response. We know that each situation is unique. However, decision-makers could consider decisions made in previous (similar) cases that consider a relation between damage in the built environment and the financial resources used to recover to help guide decision-making on reconstruction policy after a natural disaster.

## 5 Acknowledgements

The authors would like to thank the Reviewers and Editor for their helpful comments and constructive suggestions. KS received financial support from the IFTH from the Government of Ecuador

# A     Appendix 1

# B     Appendix 2

Table 1: List of articles addressing change detection with different methods and data.

| Title | Data | Method | Keywords | Year |
| --- | --- | --- | --- | --- |
| Change detection of buildings from satellite imagery and LiDAR data | Satellite images LiDAR data | Support vector machine (SVM) | building, vector, updating, change, method, database, evaluation, high, house, using | 2013 |
| Change detection of built-up land: A framework of combining pixel-based detection and object-based recognition. | high-spatial resolution remote sensing images | Differencing method, Set of morphological operations | object, pixel, change, detection, level, framework, propose, method, transformation, recognition | 2016 |
| Building change detection using old aerial images and new LiDAR data. | Satellite images Lidar data | Iterative closest point (ICP) algorithm Graph cuts method | change, building, point, method, lidar, propose, image, area, detection, cloud | 2016 |
| Object Oriented Change Detection of Buildings After a Disaster | Satellite images | Object-based change detection | building, disaster, detection, data, change, size, approach, area, discover, small | 2009 |
| Unsupervised saliency-guided SAR image change detection. | saliency-guided synthetic aperture radar (SAR) images | Saliency extraction on an initial difference map | mage, area, saliency, map, extract, obtain, difference, method, change, feature | 2017 |
| Change detection algorithm for the production of land cover change maps for countries in the European Union | high spatial resolution data | Layer arithmetic Vegetation indices Differentiating texture calculation Canonical correlation analysis (multivariate alteration detection | change, high, methodology, detection, data, classification, threshold, texture, method, map | 2014 |

Table 2: The 50-natural disasters selected for this experiment

| Country | Year | Type of Disaster | Total deaths | Total affected |
|---|---|---|---|---|
| Afghanistan | 2010 | Flood | 70 | 40000 |
| Afghanistan | 2011 | Flood | 25 | 3000 |
| Afghanistan | 2013 | Flood | 20 | 9500 |
| Afghanistan | 2014 | Flood | 431 | 140000 |
| Afghanistan | 2015 | Earthquake | | |
| Afghanistan | 2018 | Flood | 72 | 4000 |
| Bangladesh | 2018 | Flood | 21 | 14000 |
| Burkina Faso | 2010 | Flood | 16 | 133362 |
| Burundi | 2018 | Flood | | 2576 |
| Cameroon | 2015 | Flood | 4 | 30000 |
| Central African Republic | 2017 | Flood | | 3500 |
| Democratic Republic Congo | 2018 | Flood | 51 | |
| Ecuador | 2016 | Flood | 9 | 10000 |
| Haiti | 2010 | Earthquake | 222570 | $3.4*10^6$ |
| Haiti | 2016 | Flood | 13 | |
| Haiti | 2018 | Earthquake | 17 | 38915 |
| Indonesia | 2018 | Flood | | 3000 |
| Iraq | 2017 | Earthquake | 10 | 5500 |
| Kenya | 2010 | Flood | 100 | 70000 |
| Kenya | 2011 | Flood | 25 | 91692 |
| Kenya | 2012 | Flood | 73 | 280670 |
| Kenya | 2013 | Flood | 18 | 10780 |
| Libya | 2019 | Flood | 4 | 20000 |
| Mozambique | 2019 | Flood | 28 | 58000 |
| Myanmar | 2012 | Flood | 2 | 85000 |
| Myanmar | 2013 | Flood | 7 | 73300 |
| Myanmar | 2015 | Flood | 149 | $1.6217*10^6$ |
| Myanmar | 2016 | Earthquake | 4 | 1150 |
| Myanmar | 2018 | Flood | 16 | 109650 |
| Namibia | 2011 | Flood | 108 | 500000 |
| Nepal | 2015 | Earthquake | 138 | |
| Niger | 2015 | Flood | 27 | 87037 |
| Niger | 2017 | Flood | 56 | 206513 |
| Nigeria | 2017 | Flood | | 10000 |
| Nigeria | 2018 | Flood | 199 | $1.92103*10^6$ |
| Pakistan | 2010 | Flood | 46 | |
| Pakistan | 2011 | Flood | 509 | $5.4*10^6$ |
| Pakistan | 2012 | Flood | 26 | 1200 |
| Peru | 2017 | Flood | 1 | 12000 |
| Philippines | 2012 | Flood | | 25000 |
| Philippines | 2013 | Earthquake | 230 | $3.22125*10^6$ |
| Somalia | 2010 | Flood | | 16000 |
| Somalia | 2012 | Flood | 6 | 12200 |
| Somalia | 2013 | Flood | 7 | 50000 |
| South Korea | 2016 | Earthquake | | 29800 |
| South Sudan | 2012 | Flood | 32 | 154000 |
| Sri Lanka | 2010 | Flood | 28 | 606072 |
| Sri Lanka | 2011 | Flood | 18 | 225000 |